\documentstyle[amssymb,preprint,aps]{revtex}

\begin{document}
\title{The Non-mechanistic Character of Quantum Computation}
\author{Giuseppe Castagnoli, Dalida Monti}
\address{Information Communication Technology Division, Elsag spa, 16154 Genova, Italy}
\maketitle

\begin{abstract}
The higher than classical efficiency exhibited by some quantum algorithms is
here ascribed to their non-mechanistic character, which becomes evident by 
{\em joining }the notions of entanglement and quantum measurement.
Measurement analogically sets a (partial) constraint on the output of the
computation of a hard-to-reverse function. This constraint goes back in time
along the {\em reversible} computation process, computing the reverse
function, which yields quantum efficiency. The evolution, comprising wave
function collapse (here a revamped notion), is non-mechanistic as it is
driven by {\em both} an initial condition and a final constraint. It seems
that the more the output is constrained by measurement, the higher can be
the efficiency. Setting a complete constraint, by means of a special Zeno
effect, yields (speculatively) NP-complete=P.
\end{abstract}

\date{\today}

\section{Introduction}

The problem of understanding why quantum computation is more efficient than
classical computation has recently received a systematic attention: a reason
fundamental enough might be leveraged for broadening the spectrum of
efficient quantum algorithms. Ekert and Jozsa (1998) have demonstrated that
quantum entanglement is essentially involved in providing the efficiency
(see also Kitaev, 1997), but\ until now quantum measurement has been left in
the background. We will show that, by explicitly considering it, quantum
computation reveals a non-mechanistic or teleological character as far as it
is ``intelligently'' driven by {\em both} initial and final conditions.

Thus, mechanism (everything driven by initial conditions, randomness $\equiv 
$ ignorance of initial conditions, blindness to final conditions), a central
dogma of classical science, would be scientifically disproved by quantum
computation efficiency (first shown by Deutsch, 1985).

This will be shown by working on Simon's algorithm (Simon, 1994),  which is
summarized in the following. Given a function $f:B^{n}\rightarrow B^{n}$,
with $B=\{0,1\}$, 2-to-1 with periodicity $r$ (e.g. fig. 1a), the problem is
finding $r$ in poly (n) steps. We have to assume that, given a value $%
\overline{x}$ of $x$, the computation of $f\left( \overline{x}\right) $
requires poly(n) steps, whereas given a value $\overline{f}$ of $f\left(
x\right) $, the computation of $\overline{x}$ and $\overline{x}+r$ such that 
$\overline{f}=f\left( \overline{x}\right) =f\left( \overline{x}+r\right) $
requires exp(n) steps: the function must be hard to ``reverse'' with
classical computation (``invert'' is avoided since the function has no
inverse).

\bigskip

\begin{center}
(a)\qquad \qquad \qquad \qquad \qquad (b)

Fig. 1
\end{center}

Let $a$ $\left( b\right) $ be the register containing $x$ $\left( y\right) $%
, $H_{a}$ be the Hadamard transform on register $a$, $N=2^{n}$. Simon's
algorithm (fig. 1b) is implemented through the following {\em actions}%
\footnote{%
This work is influenced by the idea (due to Finkelstein, 1996) that there
are only actions, initial and final ones. } :

\noindent a) prepare: $\left| \varphi \left( t_{0}\right) \right\rangle
=\left| 0\right\rangle _{a}\left| 0\right\rangle _{b};$

\noindent b) perform $H_{a}$: $\left| \varphi \left( t_{1}\right)
\right\rangle =\frac{1}{\sqrt{N}}\sum_{x}\left| x\right\rangle _{a}\left|
0\right\rangle _{b};$

\noindent c) for each $x,$ compute $f\left( x\right) $, put result in $b:$ $%
\left| \varphi \left( t_{2}\right) \right\rangle =\frac{1}{\sqrt{N}}%
\sum_{x}\left| x\right\rangle _{a}\left| f\left( x\right) \right\rangle
_{b}; $

\noindent d) measure $f\left( x\right) $, obtaining say $\overline{f}:$ $%
\left| \beta \left( t_{3}\right) \right\rangle =\frac{1}{\sqrt{2}}\left(
\left| \overline{x}\right\rangle _{a}+\left| \overline{x}+r\right\rangle
_{a}\right) \left| \overline{f}\right\rangle _{b}$ ($\left| \beta \left(
t_{3}\right) \right\rangle $ is not a function of the former states only,
this is emphasized by changing notation from $\varphi $ to $\beta $); step
(d) is unnecessary but makes understanding easier;

\noindent e) perform $H_{a}:$ \ $\left| \beta \left( t_{4}\right)
\right\rangle =\frac{1}{\sqrt{2}}\sum_{z}\left( -1\right) ^{\overline{x}%
\cdot z}\left[ 1+\left( -1\right) ^{r\cdot z}\right] \left| z\right\rangle
_{a}\left| \overline{f}\right\rangle _{b};$ \noindent the sign $\cdot $
denotes the module 2 internal product of two binary numbers (seen as row
matrices);

\noindent f) measure $z$ (time $t_{5}$): $r\cdot z$ must be 0 for registered 
$z$; see the form of $\left| \beta \left( t_{4}\right) \right\rangle $;

\noindent g) by repeating the overall process a number of times poly(n) on
average, a number of constraints $r\cdot z^{\left( i\right) }=0$ sufficient
to identify $r$ is gathered.

\section{The notion of non-mechanistic computation}

In order to show the non-mechanistic character of Simon's algorithm, it is
convenient to circumscribe its central part from $t_{1}$ to $t_{3}$ where
``efficiency'' is achieved: the algorithm leading and trailing edges involve
neither entanglement nor quantum efficiency. To facilitate exposition, let
us say that $\left| \beta \left( t_{3}\right) \right\rangle $ already
contains the {\em readable} period $r$: readable by means of the algorithm
trailing edge, in a polynomial number of repetitions of the whole process;
or {\em readable} for short by ignoring polynomial differences of efficiency.

With reference to steps (c) and (d), we should note that $t_{2}<t_{3}$. $%
t_{2}\leqslant t_{3}$ would allow for $t_{2}=t_{3}$, making the quantum
state two-valued ($\left| \varphi \left( t_{2}\right) \right\rangle $ and $%
\left| \beta \left( t_{3}\right) \right\rangle $ at the same time), a
possibility that should be discarded. It is convenient to introduce the
notation $\left| \beta \right\rangle _{a}=\frac{1}{\sqrt{2}}\left( \left| 
\overline{x}\right\rangle _{a}+\left| \overline{x}+r\right\rangle
_{a}\right) $, thus $\left| \beta \left( t_{3}\right) \right\rangle =\left|
\beta \right\rangle _{a}\left| \overline{f}\right\rangle _{b}$. Besides
producing the {\em random} {\em outcome} $\overline{f}$ \ in register $b$,
collapse{\em \ }makes{\em \ }an{\em \ intelligent choice }in{\em \ }register 
$a$, by selecting the superposition $\left| \beta \right\rangle _{a}$ which
contains the ``readable'' period $r$, thus leading toward problem solution.
The essential point is that $\left| \beta \right\rangle _{a}$ is a
non-redundant function of both the initial condition $\left| \varphi \left(
t_{2}\right) \right\rangle $ and the measurement outcome $\overline{f}%
=f\left( t_{3}\right) $, namely a final condition occurring at time $%
t_{3}>t_{2}$ and not univocally determined by the initial condition since
wave function collapse is in between:

\begin{equation}
\left| \beta \right\rangle _{a}=\sqrt{\frac{N}{2}}\left\langle f\left(
t_{3}\right) \right| _{b}\left| \varphi \left( t_{2}\right) \right\rangle .
\end{equation}

\noindent According to equation (1), the mechanistic notion that everything
is determined by initial conditions (with randomness $\equiv $ ignorance of
initial conditions), holding in classical computation, is violated: $\left|
\beta \right\rangle _{a}$ is clearly determined by both initial and final
conditions.

We will show how non-mechanism is leveraged in Simon's algorithm from two
different perspectives:

(i) Things might be clearer by back-dating the outcome of collapse at time $%
t_{1+}$\footnote{%
According to von Neumann, collapse is atemporal as it can be back-dated any
time during the unobserved life of the quantum system between inital and
final measurement.}, after step (b) and before (c).\ The final actions of
(in reverse order) registering $\overline{f,}$ measuring register $b$, and
computing $f\left( x\right) $, change $\left| \varphi \left( t_{1}\right)
\right\rangle $ into $\left| \beta \left( t_{1+}\right) \right\rangle =\frac{%
1}{\sqrt{2}}\left( \left| \overline{x}\right\rangle _{a}+\left| \overline{x}%
+r\right\rangle _{a}\right) \left| 0\right\rangle _{b},$ where the arguments 
$\overline{x}$ and $\overline{x}+r$ are such that their function (computed
afterward) ${\em will\ be}$ $\overline{f}$; these final actions have
therefore reversed the computation of the direct function, by running it
back in time (starting from $t_{3}$ and $\overline{f}$), thus taking the
same time and achieving higher than classical efficiency.

In equivalent terms, it can be said that the measurement outcome goes back
in time yielding efficiency, it is therefore essential that the computation
process is {\em reversible} (Bennett, 1979; Fredkin and Toffoli, 1982). We
should note that, according to the current model, this backward propagation
is {\em confined} within the unobserved, reversible life of the quantum
system, namely between the initial measurement (required to prepare the
system) and the final measurement, without any possibility of carrying
information back in time in the classical world.

(ii) The following yields another perspective. The final action of measuring 
$f\left( x\right) $ both {\em creates }the output {\em \ }constraint $%
f\left( x\right) =f\left( x+r\right) =\overline{f}$ and {\em selects} the
superposition of the two arguments $\overline{x}$ and $\overline{x}+r$ which
satisfy it. Introducing and satisfying this constraint is an essential step
to solve the problem; doing both things at once, analogically, yields
quantum computation efficiency. On the contrary, of course no independent
constraint can be set on the output of {\em classical}, mechanistic
computation, which is the {\em deterministic} propagation of the input of a
reversible Boolean network.

The teleological character of quantum computation is clear: we are dealing
with a time evolution (comprising collapse) satisfying (i.e. driven by) both
an input condition and an output constraint $-$ as implied by eq. (1). We
should note that putting a constraint on\ the output of a hard to reverse
function is a general way of creating an NP problem.

Interestingly, non-mechanism brings us to re-examine the notion of causality
in the quantum framework. Causality should go forward in time during the
direct computation of $f$ and backward in time during reverse computation,
in the same time-interval. Let us see that there is no contradiction. In
fact both the forward propagation $\left| \varphi \left( t\right)
\right\rangle $ and the backward propagation $\left| \beta \left( t\right)
\right\rangle $\footnote{$\left| \beta \left( t\right) \right\rangle $
undergoes the same transformations of $\left| \varphi \left( t\right)
\right\rangle $, but with a different initial (or final) condition: while $%
\left| \varphi \left( t\right) \right\rangle $ starts from the preparation, $%
\left| \beta \left( t\right) \right\rangle $ (deterministically) ends into
the measurement outcome $\left| \beta \left( t_{3}\right) \right\rangle $,
or it goes back in time starting from that outcome.} ($\varphi $ and $\beta $
stand for forward and  backward in time) are, up to an undefined overall
phase $\delta $ (Castagnoli, 1995):

\[
\left| \varphi \left( t\right) \right\rangle =\left| \psi \left( t\right)
\right\rangle _{+}-\left| \psi \left( t\right) \right\rangle _{-},\text{ }%
\left| \beta \left( t\right) \right\rangle =e^{i\delta }\left( \left| \psi
\left( t\right) \right\rangle _{+}+\left| \psi \left( t\right) \right\rangle
_{-}\right) ,\text{ } 
\]

\noindent where $\left| \psi \left( t\right) \right\rangle _{+}$ is the {\em %
retarded wave}, associated with forward-in-time causality, and $\left| \psi
\left( t\right) \right\rangle _{-}$ is the {\em advanced wave}, associated
with backward-in-time causality, both undergoing the same transformations of 
$\left| \varphi \left( t\right) \right\rangle $. Without going into detail,
let us give such two waves in the central part of Simon's algorithm:

$\left| \psi \left( t_{1}\right) \right\rangle _{\pm }=\left| \psi \left(
t_{1+}\right) \right\rangle _{\pm }=\pm \frac{1}{2}\left[ \frac{1}{\sqrt{N}}%
\sum_{x}\left| x\right\rangle _{a}\pm \frac{1}{\sqrt{2}}e^{i\delta }\left(
\left| \overline{x}\right\rangle _{a}+\left| \overline{x}+r\right\rangle
_{a}\right) \right] \left| 0\right\rangle _{b},$

\bigskip

$\left| \psi \left( t_{2}\right) \right\rangle _{\pm }=\left| \psi \left(
t_{3}\right) \right\rangle _{\pm }=\pm \frac{1}{2}\left[ \frac{1}{\sqrt{N}}%
\sum_{x}\left| x\right\rangle _{a}\left| f\left( x\right) \right\rangle
_{b}\pm \frac{1}{\sqrt{2}}e^{i\delta }\left( \left| \overline{x}%
\right\rangle _{a}+\left| \overline{x}+r\right\rangle _{a}\right) \left| 
\overline{f}\right\rangle _{b}\right] .$

\noindent The upper (lower) signs apply to the retarded (advanced) wave. The
time-symmetry of a {\em reversible }process imposes a gauge symmetry on $%
\delta $, it must be a random variable with uniform distribution in $[0,2\pi
]$: this makes the two waves mathematically indistinguishable, either wave
can thus be associated with either direction of causality. In conclusion,
both directions of causality coexist in either $\left| \varphi \left(
t\right) \right\rangle $ or $\left| \beta \left( t\right) \right\rangle $,
and there is no privileged direction at all. Parenthetically, we should note
that each wave, representing a {\em single} direction of causality, is an
incomplete description (Castagnoli, 1995) $-$ typical of the method of
random phases (Finkelstein 1996). Whereas $\left| \varphi \left( t\right)
\right\rangle $ and $\left| \beta \left( t\right) \right\rangle $ are pure
quantum states: $\delta $ either disappears or becomes an irrelevant overall
phase.

Of course, advanced-retarded wave indistinguishability does not affect the
asymmetry between preparation and measurement outcome (thus also between $%
\left| \varphi \left( t\right) \right\rangle $ and $\left| \beta \left(
t\right) \right\rangle $), which comes from our capability of controlling
the state of the quantum object (see also Vaidman, 1998; Aharonov et al.,
1964).

While the preparation is under complete control, obtained by correcting a
previous, random measurement outcome, the final measurement outcome is not
under complete control in any of the current algorithms. Partial control
(yielding the ``intelligent choice'') is what provides quantum efficiency in
some NP\ problems. We will show in Section III that total control of the
measurement outcome would yield in principle NP-complete = P.

Eventually, let us show that skipping step (d) is indifferent; we will
follow a shortcut. Whether step (d) has been performed or skipped is {\em %
indistinguishable} to the observer of register $a$ at time $t_{5}$
(otherwise there could be superluminal communication between space-separated
regions $a$ and $b$). Therefore it is {\em equivalent} to keep step (d)
there even if it is not performed. This time, measurement of an entangled
state is involved, therefore we should not think that the result $z^{\left(
i\right) }$ (Section I)\ is already ``written'' in the state before
measurement and that measurement serves to ``read'' this result. Measurement 
{\em creates} the result, by constraining the output of $f$ computation.

\section{An alternative quantum computation paradigm}

We shall review a speculative but plausible algorithm which puts under
complete control a measurement outcome, yielding NP-complete = P
(Castagnoli, Sept. 1998). Let $f$ be a general Boolean function (fig. 2)
with constraints both on part of the input (which make the function hard to
reverse) and on the output; conventionally, the output constraint is 1. The
problem is whether there is an assignment of the unconstrained part of the
input $x_{1}$, $x_{2}$, ... satisfying all input and output constraints.
This is a version of the well-known NP-complete SAT problem.

\begin{center}
Fig. 2
\end{center}

The computer register must comprise both the input and output qubits
appearing in fig. 2; each unconstrained input is prepared in $\frac{1}{\sqrt{%
2}}\left( \left| 0\right\rangle +\left| 1\right\rangle \right) $. By using
conventional quantum computation, we compute the direct function, reaching
in polynomial time an entangled state $\left| \varphi \right\rangle $ which
is an equally weighted superposition of all tensor products of qubits
eigenstates which satisfy the input constraint and $f$ (while the output
constraint may not be satisfied). At this time interaction has ceased and
all registers' qubits are independent of each other.

To simplify, say that the network has exactly one solution: the problem is
to find it. All $\left| \varphi \right\rangle $'s tensor products but one
end with $\left| 0\right\rangle _{y}$. A $\pi /2$ rotation is now applied to
the independent qubit $y$ (in order to bring it from about $\left|
0\right\rangle _{y}\left\langle 0\right| _{y}$ to about $\left|
1\right\rangle _{y}\left\langle 1\right| _{y}$), while the overall network
state is (speculatively) submitted to a {\em continuous} measurement, such
that it is continuously projected on the constrained Hilbert subspace ${\cal %
H}^{c}$. By definition, ${\cal H}^{c}$ is the span of all the vectors which
satisfy the input constraints and $f$ (note that $\left| \varphi
\right\rangle $ as all subsequent measurement outcomes belong to ${\cal H}%
^{c}$). \noindent This generates a unitary propagation of the network state
which, in one computation step (the $\pi /2$ rotation), leads to a state
close to the solution (ending with $\left| 1\right\rangle _{y}$). Qubit
measurement gives the solution with high probability.

The notion of continuous measurement, as the associated ``special'' Zeno
effect (keeping the propagation inside ${\cal H}^{c}$), is strictly based on
a retarded-advanced wave model. Without it, we would remain with the
mechanistic notion of frequent measurement in the limit of infinite
frequency. This would bring in a different Zeno effect freezing the
propagation (no more driven by both initial and final conditions) in its
initial state. By the way, this discrepancy between continuous and frequent
measurement might lead to ascertaining the existence of an advanced
propagation.

\section{Discussion}

We have shown that a quantum propagation comprising measurement and wave
function collapse (here a revamped notion) is strictly non-mechanistic in
character, being driven by both initial and final conditions.

Although non-mechanism justifies the efficiency of quantum computation, it
should be noted that it came out in the past in several forms, apparently
without being acknowledged. Maybe, because the notion of causality going
back and forth in time is often  ``aborred'', considered with either no
consequences (an idle interpretation) or too many (time-travel). Hopefully,
this work should convince the reader that both concerns are not justified.

Let us see how the notion of non-mechanism (tacitly) has developed in the
past. It\ has been a progressive development, characterized by an increasing
control of the measurement outcome:

(i) In the simplest case of an elementary (un-compound) object, eq. (1)
tells that the outcome of measurement is a function of itself (is what it
is): the only constraint is that it is one of the outcomes allowed by the
state before measurement. This is freedom from the past $-$ namely
non-mechanism. Here the measurement outcome can be controlled only in a
stochastic way (the probability amplitudes can be prepared).

(ii) In the case of a compound object in an entangled state, non-mechanism
also appears in the novel form of {\em correlation} between simultaneous
eigenvalues of the measurement outcome $-$ like in EPR measurement. This is
called non locality of course. However, such a correlation can be described
without giving up locality, provided that causality is allowed to go back
and forth in time (Bennett, 1995; Castagnoli, 1995).

(iii) A third level of unfolding of non-mechanism appears when it is
leveraged to yield quantum computation efficiency. In current algorithms, as
already in (ii), part of the outcome is random, part is correlated. This
yields a capability of solving in polynomial time some NP problems. We have
shown a further level of unfolding of non-mechanism, speculative for the
time \ being. This corresponds to putting the outcome of measurement under a
complete control, by means of a special Zeno effect. This would allow to
solve the SAT\ problem in polynomial time, in principle.

We should further note that, through steps (i)-(iii), the teleological
character of the evolution, more and more controlled by the final condition,
increases.

Non-mechanism appears to be an exclusively quantum feature of evolutions
comprising wave function collapse. Like all exclusively quantum features, on
the one hand it is strictly confined between initial and final measurement,
on the other hand it yields consequences in the classical world, in fact
quantum computation efficiency.

Hopefully, the perspective developed in this work will help in the quest for
new efficient computation algorithms. Thanks are due to G. Baget Bozzo, T.
Beth, A. Ekert, D. Finkelstein, and V. Vedral for stimulating discussions.

\bigskip

{\LARGE References}

\begin{enumerate}
\item  Y. Aharonov, P.G. Bergmann, and J.L. Lebowitz (1964), {\em Phys. Rev.}
{\bf 134} {\bf B}, 1410.

\item  C.H. Bennett (1979), Logical Reversibility of Computation, {\em IBM
J. Res. Dev.} {\bf 6}, 525.

\item  C.H. Bennett (1995), {\em Elsag Bailey-ISI\ Turin Workshop on Quantum
Computation and Communication}, oral discussion.

\item  G. Castagnoli (September 1998), NP-complete=P under the Hypothetical
Notion of Continuous von Neumann Measurement, {\em Helsinki QCC Pathfinder
Conference}; quant-ph/9810017.

\item  G. Castagnoli (1995), {\em Int. J. Theor. Phys.}, {\bf 34}, 1283.

\item  G. Castagnoli (1998), {\em Phys.D} {\bf 120}, 48; quant-ph/9706019.

\item  G. Castagnoli and Dalida Monti (1998), Quantum Computation Based on
Particle Statistics, to appear in {\em Chaos, Solitons and Fractals};
quant-ph/9806010.

\item  D. Deutsch (1985), {\em Proc. Roy. Soc. London} {\bf A 400}, 97.

\item  A. Ekert, R. Jozsa (1998), Quantum Algorithm: Entanglement Enhanced
Information Processing, to appear in {\em Phil. Trans. Roy. Soc.} {\em %
(Lond.)}; quant-ph/9803072.

\item  D.R. Finkelstein (1996), Quantum Relativity, {\em Springer, Berlin
Heidelberg}.

\item  E. Fredkin and T. Toffoli (1982),{\em \ Int. J. Theor. Phys.} {\bf 21}%
, 219.

\item  A.Y. Kitaev (1997), Fault-Tolerant Quantum Computation By Anyons,
manuscript: quant-ph/9707021.

\item  D.R. Simon (1994), {\em Proceedings of the 35}$^{{\em th}}${\em \
Annual Symposium on the Foundation of Computer Science},{\em \ }Santa Fe,
IVM.

\item  L. Vaidman (1998), Time-Symmetrized Counterfactuals in Quantum
Theory, manuscript.
\end{enumerate}

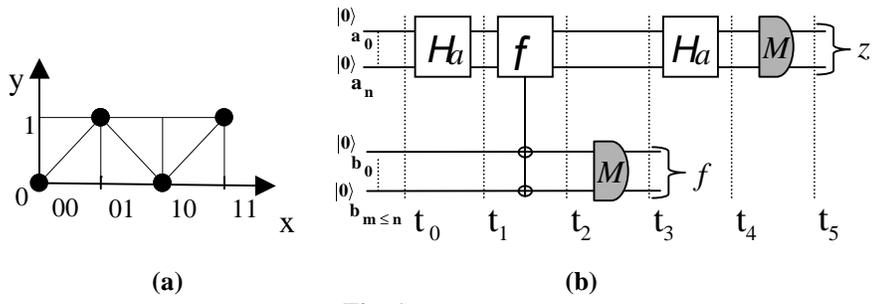
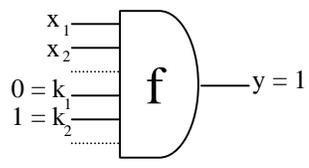
\end{document}